\documentclass[%
 reprint,
 amsmath,
 amssymb,
 aip,
]{revtex4-2}

\usepackage{listings}
\usepackage{graphicx}
\usepackage{dcolumn}
\usepackage{bm}
\usepackage{multirow}
\usepackage{siunitx} 
\usepackage[hidelinks]{hyperref}

\usepackage[normalem]{ulem} 

\usepackage{tikz}

\newcommand{\SIdeltas}{S1}
\newcommand{\SIposdepNESS}{S1}
\newcommand{\SIlipid}{S2}
\newcommand{\SImethods}{S2}
\newcommand{\SIumbrella}{S3}
\newcommand{\SIhyper}{S4}
\newcommand{\SIalgo}{S1}
\newcommand{\subref}[2]{\hyperref[#1]{\ref*{#1}#2}} 
\newcommand{\pness}{p^\mathrm{ss}}

\begin{document}

\title{Fokker--Planck Score Learning: Efficient Free-Energy Estimation under Periodic Boundary Conditions}

\author{Daniel Nagel}
\affiliation{%
 Institute for Theoretical Physics, Heidelberg University, 69120 Heidelberg, Germany
}%
\author{Tristan Bereau}%
\email{bereau@uni-heidelberg.de}
\affiliation{%
 Institute for Theoretical Physics, Heidelberg University, 69120 Heidelberg, Germany
}%
\affiliation{%
 Interdisciplinary Center for Scientific Computing (IWR), Heidelberg University, 69120 Heidelberg, Germany
}%

\date{1 October 2025}

\begin{abstract}
Accurate free-energy estimation is essential in molecular simulation, yet the
periodic boundary conditions (PBC) commonly used in computer simulations have
rarely been explicitly exploited. Equilibrium methods such as umbrella sampling,
metadynamics, and adaptive biasing force require extensive sampling, while
non-equilibrium pulling with Jarzynski's equality suffers from poor convergence
due to exponential averaging. Here, we introduce a physics-informed, score-based
diffusion framework: by mapping PBC simulations onto a Brownian particle in a
periodic potential, we derive the Fokker--Planck steady-state score that
directly encodes free-energy gradients. A neural network is trained on
non-equilibrium trajectories to learn this score, providing a principled scheme
to efficiently reconstruct the potential of mean force. On benchmark
periodic potentials and small-molecule membrane permeation, our method is up to
one order of magnitude more efficient than umbrella sampling.
\end{abstract}

\maketitle

\section{Introduction}

Free-energy profiles quantify the thermodynamic stability and transition
kinetics of molecular states, underpinning predictions in protein folding,
ligand binding, and materials design.\cite{chipot2007free, hansen2014practical,
frenkel2023understanding} Estimating these profiles typically relies on
particle-based molecular simulations---most commonly molecular dynamics---to
sample the relevant high-dimensional configuration space. Modern simulations
almost universally employ periodic boundary conditions (PBC) to mimic bulk
environments and eliminate surface artifacts, which leads to repeating energy
landscapes across unit cells that must be navigated by any free-energy
estimator. Interestingly, none of the conventional free-energy estimation
methods explicitly exploit periodicity.

Equilibrium sampling approaches remain foundational for free-energy
calculations.  Metadynamics adaptively deposits history-dependent bias to escape
deep minima,\cite{laio2002escaping} adaptive biasing force (ABF) methods apply
on-the-fly forces to flatten barriers,\cite{darve2008adaptive} while other
methods, such as replica-exchange\cite{sugita1999replica} and accelerated
molecular dynamics,\cite{hamelberg2004accelerated} can further enhance sampling
efficiency. Nevertheless, umbrella sampling--based methods remain the current
gold standard: by applying overlapping harmonic restraints, umbrella
sampling\cite{torrie1977nonphysical, roux1995calculation}
ensures comprehensive coverage of high-energy regions, and the
weighted histogram analysis method (WHAM)\cite{kumar92, hub10} or multistate Bennett
acceptance ratio (MBAR)\cite{shirts2008} estimators reconstruct the
unbiased potential of mean force (PMF) from the biased samples.
Umbrella sampling--based methods have become a workhorse method to estimate
PMFs for various biomolecular applications, including protein-ligand
binding,\cite{orsi2010passive, buch2011optimized, swift2013back} atomistic
simulations of drugs in phospholipid bilayers, \cite{carpenter2014method,
lee2016simulation, bennion2017predicting, tse2018link} and high-throughput
screening of solute permeation through lipid
bilayers.\cite{menichetti2017silico, menichetti2019drug}

Non-equilibrium pulling offers a complementary route by mechanically driving
systems along reaction coordinates and employing Jarzynski's equality to relate
work distributions to equilibrium free-energy
differences.\cite{jarzynski1997nonequilibrium} In practice, however, the
requisite exponential averaging converges poorly unless work fluctuations are
tightly controlled, as demonstrated in helix-coil transitions,\cite{park04}
ion dissociation,\cite{wolf18} ion-channel ligand unbinding,\cite{bacstuug2008potential} and solute permeation
through lipid bilayers.\cite{noh2020comparison}
Taken together, Jarzynski-based estimators struggle to converge as efficiently
as equilibrium methods.
These challenges underscore the
need for stronger, physics-informed inductive biases to guide free-energy
inference from non-equilibrium data. Our approach builds on recent
neural-sampler methods\cite{noe2019boltzmann, wirnsberger2020targeted,
wirnsberger2023estimating, mate24, mate2025solvation, he2025feat} and is
complementary to emerging diffusion-based strategies that interpolate
thermodynamic observables or accelerate enhanced
sampling\cite{herron2024inferring, obi2024application, benayad2025hamiltonian},
but departs by explicitly exploiting PBC and embedding the analytic
nonequilibrium periodic steady state directly in the score.

\begin{figure*}
    \centering
    \includegraphics{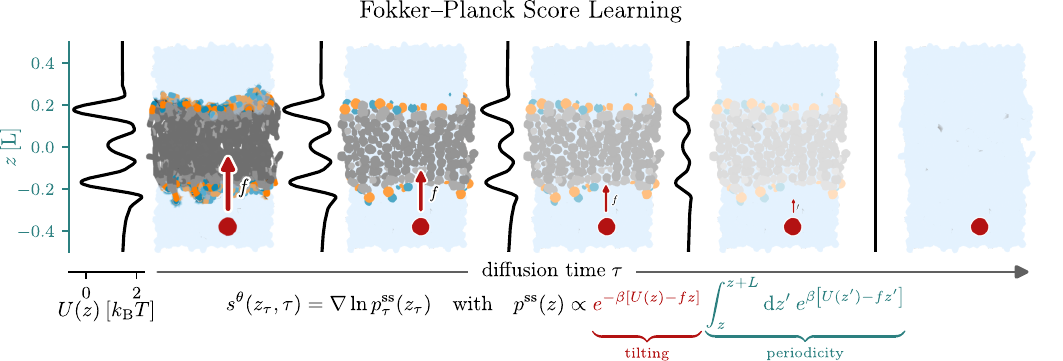}
    \caption{We consider a computer simulation with periodic boundary
    conditions, where the red particle is pulled by a constant external driving
    force $f$ through a conservative potential or potential of mean force, $U(x)$.
    The steady-state solution of the Fokker--Planck equation of a Brownian
    particle in a periodic potential, $\pness$, informs the score of our
    diffusion model. The diffusion model interpolates between the original
    non-equilibrium system at constant flux $J$ and a trivial system at rest,
    shown left to right, respectively. Denoising of the diffusion model allows
    us to efficiently reconstruct $U(x)$ by exploiting the structure of
    $\pness$.}
    \label{fig:method}
\end{figure*}

Most particle-based computer simulations employ
PBC to mimic bulk environments and eliminate edge effects. This introduces a
repeating energy landscape akin to multiple identical barriers in series, a
scenario naturally captured by the classical model of a Brownian particle
diffusing in a periodic potential.
Consider a particle moving in a one-dimensional periodic potential under a
constant external driving force $f$. This prototypical non-equilibrium steady
state has the analytical probability density
\begin{equation}
    \label{eq:fp_steady_state}
    \pness(x) \propto e^{-\beta [U(x) - f x]}\, \int_{x}^{x+L} \mathrm{d}y\, e^{\beta [U(y) - f y]}\,,
\end{equation}
where $U(x) - f x$ is the tilted effective potential function, $\beta =
(k_\mathrm{B}T)^{-1}$ the inverse temperature, and $L$ the spatial
period.\cite{risken1996fokker} The first term on the right-hand side is the
local Boltzmann factor of the tilted potential. In the absence of PBCs the
steady-state distribution would simply be this tilted equilibrium form.
However, enforcing periodicity introduces the second, nonlocal period-integral
term, which is essential to guarantee the constant probability current of the
Brownian particle.\cite{reimann2002diffusion, ma2015colloidal}
This expression links the non-equilibrium steady-state
distribution to the underlying equilibrium potential.
We adopt it as an ansatz for parameterizing our diffusion model, leveraging its
powerful inductive bias to reconstruct free-energy landscapes in periodic systems.

Building on this mapping, we employ a score-based diffusion model to learn the
so-called Fokker--Planck score, which is parametrized by the steady-state
solution of the Fokker--Planck equation for our periodic system.
The original data distribution
corresponds to the non-equilibrium setup at finite flux, $J>0$, while the
trivial latent distribution turns off both conservative and non-equilibrium
effects to yield $J=0$, see Fig.~\ref{fig:method}.  Rather than the flux itself, we parametrize a time interpolation of the combined energy function and pulling force, and, if applicable, the spatially-dependent diffusion coefficient.  The diffusion model learns
to interpolate between these two distributions via its score, $s(x,\tau) = \nabla
\ln p_\tau(x)$, where $p_\tau(x)$ is the diffusion-time dependent density of the
diffusion model.\cite{hyvarinen2005estimation, song2019generative,
song2020score} When the system is in equilibrium, the score can be informed by
directly fitting the (diffusion-time-dependent) force, $s(x,\tau) = - \beta \nabla
U_\tau(x)$.\cite{arts2023two, mate24, mate2025solvation} Here instead, we inform
$p_\tau(x)$---and thus the score---through the functional form of
Eq.~\eqref{eq:fp_steady_state}. Training of our model follows a denoising
score-matching approach.\cite{vincent2011connection} By training a neural
network on non-equilibrium trajectory data, we recover the steady-state score
across the periodic box and integrate it to reconstruct the underlying PMF. We 
coin this approach ``Fokker--Planck Score Learning'', as a principled
route to reconstructing free-energy landscapes of periodic systems.  We will
demonstrate that the inductive bias provided by Fokker--Planck Score Learning is
particularly advantageous in the low-data regime, where conventional methods
struggle to converge.

The remainder of this paper is organized as follows. In
section \ref{sec:theory}, we recall the derivation of the Brownian particle in a
periodic potential, obtain the Fokker--Planck steady state, and propose the
resulting score. Section \ref{sec:results} demonstrates the performance and
versatility of our proposed method, where we apply it to two representative test
cases: ($i$) a one-dimensional toy potential and ($ii$) the insertion of a
solute in a phospholipid bilayer modeled with the coarse-grained Martini 3 force
field. In each case, we reconstruct the underlying free-energy landscape via a
continuous-time score-based diffusion model together with the analytic
steady-state solution of the Fokker--Planck equation for periodic,
nonequilibrium systems. We show that our approach achieves faster convergence
and lower variance compared to conventional umbrella sampling analyzed with MBAR and WHAM.

\section{Theory and Methods\label{sec:theory}}

\subsection{Nonequilibrium Steady-States}
This section briefly outlines the derivation of the steady-state
distribution for a Brownian particle in a periodic potential. This distribution
forms the inductive bias for our proposed diffusion model. For a comprehensive
derivation, readers are referred to standard texts, such as
Risken.\cite{risken1996fokker}

Consider a particle governed by overdamped Langevin dynamics
\begin{equation}
    \frac{\textrm{d}x}{\textrm{d}t} = -\beta D \nabla U(x) + \sqrt{2D}\xi(t)\,,
\end{equation}
where $\xi(t)$ represents Gaussian white noise with $\langle\xi(t)\rangle = 0$
and $\langle \xi(t)\xi(t')\rangle = \delta(t-t')$, $D$ is the diffusion coefficient,
$\beta$ denotes the inverse temperature, and $U(x)$ is the
potential energy function. It is assumed that the Einstein relation, $D\beta =
\mu$, is valid, where $\mu$ is the particle mobility.
The corresponding Fokker--Planck equation for this system is
\begin{equation}\label{eq:fokker_planck}
    \partial_t p_t(x) = \nabla\left\{ \beta D \nabla U(x)\, p_t(x)\right\} + D\,\nabla^2p_t(x)\,,
\end{equation}
where $p_t(x)$ is the probability density at time $t$. The stationary solution
to this equation is the Boltzmann distribution
\begin{equation}
    \pness(x) \propto e^{-\beta U(x)}\,.
\end{equation}
When a constant external driving force, $f(x)=f$, is introduced, the
overdamped Langevin equation is modified to
\begin{equation}
    \frac{\textrm{d}x}{\textrm{d}t} = -\beta D\left[ \nabla U(x) - f\right] + \sqrt{2D}\xi(t)\,.
\end{equation}
Consequently, an effective potential energy function can be defined as
\begin{equation}
    U_\textrm{eff}(x) = U(x) - fx\,. 
\end{equation}
The non-equilibrium steady-state (NESS) distribution then reflects the
combined influence of the potential and the external force
\begin{equation}
\pness(x) \propto e^{-\beta U_\textrm{eff}(x)} \,.\label{eq:ness_nopbc}
\end{equation}

It is important to note that, as demonstrated by Koch \emph{et al.},\cite{koch25} the
derivation of a Langevin equation using projection operator formalism reveals
that an external force applied to a particle does not, in general, enter
additively into the resulting equation. However, this additive approximation is
valid under the assumption of weak driving forces, which is adopted in this
work.

\subsection{Periodic Systems}

We now extend the analysis to driven systems subject to periodic boundary
conditions. In such systems, the configuration variable $x$ is defined on a
domain of period $L$, such that $x \in [0,L]$. Consequently, all physical
observables $\mathcal{O}$ are periodic functions of $x$, satisfying
\begin{equation}
\mathcal{O}(x+L)=\mathcal{O}(x)\,.
\end{equation}
This implies that the potential energy function $U(x)$ must also be periodic,
$U(x+L) = U(x)$. The periodicity naturally extends to the non-equilibrium
steady-state (NESS) distribution, $\pness(x)$. Therefore,
$\pness(x)$ cannot be described by a simple tilt of the potential
energy, as was the case in Eq.~\eqref{eq:ness_nopbc} for non-periodic systems.

For a driven periodic system, the Fokker--Planck equation governing the time
evolution of the probability distribution $p_t(x)$ retains its previous form
\begin{align}
    \partial_t p_t(x) &= \nabla\left\{ \beta D \nabla U_\textrm{eff}(x)\, p_t(x)\right\} + D\,\nabla^2p_t(x)\\
    &= - \nabla J_t(x)\,,
\end{align}
where $U_\textrm{eff}(x) = U(x) - fx$ is the effective potential energy, as
defined earlier, and $J_t(x)$ denotes the probability current density
\begin{equation}
    J_t(x) = -\beta D \nabla U_\textrm{eff}(x)\, p_t(x) - D\,\nabla p_t(x)\,.
\end{equation}

In the steady-state limit, the probability current $J_t(x)=J$ becomes constant.
By introducing an auxiliary function $\kappa(x) = e^{\beta U_\textrm{eff}(x)}$, the
Fokker--Planck equation for the NESS distribution can
be rewritten as
\begin{equation}
    \nabla (\kappa(x) \pness(x)) = - \frac{J}{D} \kappa(x)\;.
\end{equation}
Integrating this equation over one period, from $x$ to $x + L$, and utilizing
the periodicity conditions $\pness(x) = \pness(x+L)$ and
$\kappa(x+L) = \kappa(x) e^{-\beta fL}$, yields the NESS distribution
\begin{align}
    \pness(x) 
    &\propto e^{-\beta U_\text{eff}(x)}\, \int_{x}^{x+L} \textrm{d}y\: e^{\beta U_\text{eff}(y)}\,.
\end{align}

This result can be generalized to accommodate a position-dependent diffusion
coefficient $D(x)$. As detailed in the Supporting Information (see section
\SIposdepNESS), this generalization leads to a modified expression for the NESS
distribution
\begin{equation}
    \label{eq:ness_pbc}
    \pness(x) \propto \frac{1}{D(x)} e^{-\beta U_\text{eff}(x)}\, \int_{x}^{x+L} \textrm{d}y\: e^{\beta U_\text{eff}(y)}\,.
\end{equation}

\subsection{Denoising Diffusion Models on Periodic Domains}

\subparagraph{Denoising Diffusion Models}
Denoising diffusion models (DDMs) represent a class of generative models that
learn a target data distribution, $p_{\text{data}}(x)$, by systematically
reversing a predefined, gradual noising
process.\cite{sohldickstein15,ho20,song2020score} While various noising schemes can be
employed, we focus here on an overdamped Langevin process.
The forward process describes a relaxation of the data distribution in the
dimensionless prior potential $\phi(x)$, for the diffusion time $\tau \in [0, 1]$
according to the stochastic differential equation
(SDE)
\begin{equation}
    \mathrm{d}x_\tau = -\alpha_\tau\nabla \phi(x)\mathrm{d}\tau + \sqrt{2\alpha_\tau}\mathrm{d}W_\tau\,,
    \label{eq:ddm_fwd_sde_general}
\end{equation}
where $\alpha_\tau > 0$ is a time-dependent noise schedule, $\phi(x)$ is a
dimensionless potential defining the latent prior distribution (often chosen to
be harmonic), and $W_\tau$ denotes a standard Wiener process.

The corresponding Fokker--Planck equation governs the evolution of the marginal
probability density $p_\tau(x)$ at diffusion time $\tau$\cite{saerkkae19}
\begin{equation}
    \partial_\tau p_\tau(x) = \alpha_\tau\nabla\left[\nabla\phi(x) p_\tau(x) + \nabla p_\tau(x)\right]\,.
    \label{eq:ddm_fwd_fp_general}
\end{equation}
By defining a physical time $t$ such that $D\mathrm{d}t =
\alpha_\tau \mathrm{d}\tau$, and setting $\phi(x) = \beta U(x)$,
Eq.~\eqref{eq:ddm_fwd_fp_general} becomes equivalent to the standard
Fokker--Planck equation (cf. Eq.~\eqref{eq:fokker_planck}) for a system with
potential $U(x)$ and diffusion coefficient $D$. In the DDM context, however,
$\alpha_\tau$ is primarily a design choice optimized for model training and may not
directly correspond to a physical diffusion rate.

The generative, or reverse, process aims to invert this noising procedure. The
reverse-time SDE, corresponding to Eq.~\eqref{eq:ddm_fwd_sde_general}, is given
by\cite{anderson82}
\begin{equation}
    \label{eq:ddm_rev_sde}
    \mathrm{d}x_\tau = -\alpha_\tau\nabla\left[\phi(x) + 2\ln p_\tau(x_\tau)\right] \mathrm{d}\tau + \sqrt{2\alpha_\tau}\mathrm{d}\bar{W}_\tau\,.
\end{equation}
Here, $\mathrm{d}\tau$ represents a negative infinitesimal time step (i.e., the
process evolves from $\tau=1$ to $\tau=0$), and $\bar{W}_\tau$ is a standard Wiener
process under reverse time flow. The crucial component for this reverse process
is the score function, $s(x_\tau,\tau) = \nabla \ln p_\tau(x_\tau)$. By learning an
approximation to this score function, typically with a neural network
$s_\theta(x_\tau,\tau)$, and solving Eq.~\eqref{eq:ddm_rev_sde} numerically, samples
from the prior distribution $p_1(x)$ (associated with $\phi(x)$ at $\tau=1$) can be
transformed into samples approximating the target data distribution $p_0(x)
\approx p_{\text{data}}(x)$.

\subparagraph{Learning the Score}
The score function $s_\theta(x_\tau,\tau)$ is learned by minimizing the
following objective function, which is derived from the variational lower bound
on the data log-likelihood\cite{song2020score}
\begin{equation}
    \mathcal{L}(\theta) = \mathbb{E}_{x_0\sim p_0, x_\tau\sim p(x_\tau|x_0)} \left[ \left\| s_\theta(x_\tau, \tau) - \nabla \ln p_\tau(x_\tau|x_0) \right\|^2 \right]\,.
\end{equation}
Using the fact that the conditional probability distribution
$p_\tau(x_\tau|x_0)$ is Gaussian, together with the solution to the forward SDE
$x_\tau = \gamma_\tau x_0 + \sigma_\tau \varepsilon$, where $\varepsilon \sim
\mathcal{N}(0,1)$ is a standard normal random variable, the target score in the objective can be expressed as
\begin{equation}
    \nabla \ln p_\tau(x_\tau|x_0) = -\frac{1}{\sigma_\tau^2} (x_\tau - \gamma_\tau x_0) = -\frac{\varepsilon}{\sigma_\tau}\,,
\end{equation}
where $\gamma_\tau$ is the drift term introduced by the prior potential $\phi(x)$
and $\sigma_\tau$ is the standard deviation of the noise at time $\tau$. Hence,
typically a neural network, $\varepsilon^\theta(x_\tau, \tau)$, is trained to predict the
noise component $\varepsilon$ given the noised sample $x_\tau$ and diffusion time
$\tau$, instead of learning the score $s(x_\tau, \tau)$ directly.

\subparagraph{Periodic Diffusion Models}
To apply DDMs to systems with periodic boundary conditions of period $L$, a
uniform distribution over the domain $[0, L)$ is chosen as the prior. This
corresponds to a constant prior potential, $\phi(x) = \ln L$, so that
$p(x) = e^{-\phi(x)} = 1/ L$, which
implies $\nabla\phi(x) = 0$. Consequently, the forward and reverse SDEs simplify
to
\begin{align}
\mathrm{d}x_\tau &= \sqrt{2\alpha_\tau}\mathrm{d}W_\tau \label{eq:ddm_fwd_sde_periodic}\\
\mathrm{d}x_\tau &= -2\alpha_\tau\nabla \ln p_\tau(x_\tau)\mathrm{d}\tau + \sqrt{2\alpha_\tau}\mathrm{d}\bar{W}_\tau\,. \label{eq:ddm_rev_sde_periodic}
\end{align}
As demonstrated by Máté \emph{et al.},\cite{mate24, mate2025solvation}
learning the noise $\varepsilon^\theta(x_\tau, \tau)$ in the forward process can be
directly extended to periodic domains by applying the modulo operation to the
diffused positions during training, i.e., $x_\tau \to (x_\tau \pmod L)$.

\subparagraph{Energy-Based Diffusion Models}
The score function, $s(x_\tau, \tau) = \nabla \ln p_\tau(x_\tau)$, is pivotal in
score-based generative models as it dictates the drift term in the reverse-time
stochastic differential equation for sample generation. However, a neural
network $s^\theta(x_\tau, \tau)$ trained to approximate this score is not
intrinsically guaranteed to be a conservative vector field, meaning it may not
identify to the gradient of a scalar potential. To enforce this conservative
property, which aligns the model with physical energy landscapes, energy-based
diffusion models are utilized.\cite{arts2023two, mate24, mate2025solvation} There, the score is explicitly
parameterized as the gradient of a time-dependent potential energy
function $U^\theta_\tau(x_\tau, \tau)$:
\begin{equation}
s^\theta(x_\tau, \tau) = -\beta \nabla U^\theta(x_\tau, \tau)\,.
\end{equation}
The gradient $\nabla U^\theta(x_\tau, \tau)$ is computed using automatic
differentiation. This methodology ensures by construction that the learned score
field is conservative, thereby directly connecting the generative process to an
underlying, learnable energy function.

\subsection{Fokker--Planck Score Learning}

To integrate the NESS simulation data with the
denoising diffusion model, we define an effective potential that interpolates
between the physical system and the prior distribution over the diffusion time
$\tau \in [0, 1]$. This time-dependent effective potential is given by
\begin{equation}\label{eq:U_eff_dm}
U_\text{eff}(x, \tau) = (1-\tau) \left[U_\tau(x) - fx\right] + \tau \ln L\,,
\end{equation}
where $U_\tau(x)$ represents the potential energy at diffusion time $\tau$, $f$ is the
constant external driving force, and $\ln
L$ corresponds to the potential of a uniform prior distribution over the
periodic domain of length $L$. This formulation employs a physical prior
regularization approach, where the physical potential $U_\tau(x)$ and the driving
term gradually diminish as $\tau \to 1$, while the uniform prior potential is
concurrently introduced.

Similarly, we define a time-dependent, normalized diffusion coefficient as
\begin{equation}
    D(x, \tau) = (1-\tau) \frac{D(x)}{\langle D \rangle_x} + \tau\,,
\end{equation}
where $D(x)$ is the position-dependent physical diffusion coefficient, and
$\langle D \rangle_x$ denotes its spatial average over the periodic domain. This
ensures that $D(x,\tau)$ transitions from the scaled physical diffusion coefficient
at $\tau=0$ to a uniform unit diffusion coefficient at $\tau=1$.

The core objective is to train a neural network $U^\theta_\tau(x)$ to approximate
the time-dependent potential $U_\tau(x)$, such that at the initial diffusion time
($\tau=0$), $U^\theta_0(x)$ provides an accurate estimate of the physical potential
$U(x)$. Leveraging the NESS distribution from Eq.~\eqref{eq:ness_pbc} and the
principles of an energy-based DDM on a periodic bound, we express the score
function as
\begin{align}
    s^\theta(x_\tau, \tau, L) =&\nabla \ln \pness(x_\tau, \tau, L)\nonumber\\
    =&-\beta \nabla U^\theta_\text{eff}(x_\tau, \tau) - \nabla \ln D(x_\tau, \tau) \nonumber\\
    &+ \Delta s^\theta(x_\tau, \tau, L)\,, \label{eq:score_with_correction}
\end{align}
where $U^\theta_\text{eff}(x_\tau, \tau)$ is the neural network approximation of the
effective potential
\begin{equation}
    U_\text{eff}^\theta(x, \tau) = (1-\tau) \left[U^\theta_\tau(x) - fx\right] + \tau \ln L\,.
\end{equation}
The term $\Delta s^\theta(x_\tau, \tau, L)$ arises from the
periodicity constraint on $\pness(x)$ and is given by
\begin{align}
    \Delta s^\theta(x_\tau, \tau, L) &= \nabla \ln \int_{x_\tau}^{x_\tau + L} \textrm{d}y\: e^{\beta U^\theta_\text{eff}(y, \tau)}\nonumber\\
    &= \frac{e^{\beta U^\theta_\text{eff}(x_\tau, \tau)}\left[e^{-\beta(1-\tau)fL} - 1\right]}%
    {\int_{x_\tau}^{x_\tau+L} \textrm{d}y\: e^{\beta U^\theta_\text{eff}(y, \tau)}}\,. \label{eq:pbc_correction_term}
\end{align}

%

\subsection{Simulation Setup and Network Parame\-trization}
\subparagraph{Molecular Dynamics Simulations}
We employed the Martini 3 coarse-grained
force field,\cite{souza21} together with the automated 
workflow Martignac for lipid-bilayer simulations.\cite{martignac24}  A symmetric
1-palmitoyl-2-oleoyl-$sn$-glycero-3-phosphocholine (POPC) bilayer was constructed,
interacting with a solute modeled by a single Martini N1 bead. All molecular dynamics
simulations were carried out with GROMACS 2024.3,\cite{berendsen1995gromacs} adhering to
the protocol of Martignac with an integration time step of $\delta t=
\SI{0.02}{\pico\second}$.\cite{martignac24}
Here and throughout this work, time scales for the Martini 3 coarse-grained
model are expressed in picoseconds.
We used a box size of $L_x = L_y = \SI{6}{\nano\meter}$ and $L_z =
\SI{10}{\nano\meter}$, with periodic
boundary conditions in all three dimensions. Each run was equilibrated for
$\SI{200}{\pico\second}$ in the isothermal-isobaric ($NPT$) ensemble at
$T=\SI{298}{\kelvin}$ and $P=\SI{1}{\bar}$, using a stochastic
velocity-rescaling thermostat\cite{bussi07} and a C-rescale
barostat\cite{bernetti20}, respectively. A semi-isotropic pressure coupling
scheme was employed, with the compressibility in the $z$-direction set to zero
to maintain a constant box height. Subsequent production runs were conducted in
the canonical ($NVT$) ensemble sharing the same temperature and thermostat.

\subparagraph{Constant Force Pulling}
We performed five independent production runs, each of $\SI{1}{\micro\second}$
duration, applying constant biasing forces $f \in \{0, 4, 6, 8,
10\}\:\si{\kilo\joule\per\mol\per\nano\meter}$. Trajectory frames were recorded
every $\delta t = \SI{0.2}{\pico\second}$ for subsequent analysis.

\subparagraph{Umbrella Sampling}
As a reference, we carried out umbrella sampling\cite{torrie1977nonphysical} over windows of width $\delta z
= \SI{0.02}{\nano\meter}$ covering the range $z \in [0, 4.8] \, \si{\nano\meter}$, with \SI{500}{\nano\second} of simulation per window.
For direct comparison with our approach, we also performed umbrella sampling at
the conventional spatial resolution of $\delta z = \SI{0.1}{\nano\meter}$, again
using \SI{500}{\nano\second} of MD per window. To reconstruct the free-energy 
profile and estimate uncertainties, the
weighted histogram analysis method (WHAM)\cite{kumar92, hub10} and multistate
Bennett acceptance ratio (MBAR)\cite{shirts2008} were employed, respectively. Both
applied to trajectory segments in the interval $t \in [1, 100]\,
\si{\nano\second}$, based on up to 50 independent replicates.

\subparagraph{Position-Dependent Diffusion Coefficient}
Various approaches exist to extract the position-dependent diffusion
coefficient $D(z)$ from MD data---for example, the umbrella-sampling based
method of Hummer\cite{hummer05}. Here, we instead determined $D(z)$ directly
from the distribution of particle displacements along the $z$-axis,
\begin{equation}
\Delta z = z(t + \Delta t) - z(t)\,,
\end{equation}
where $z(t)$ denoted the particle's position at time $t$ and $\Delta t$ was the
sampling interval.  The local diffusion coefficient was then obtained from the
variance of these displacements via
\begin{equation}
D(z) = \frac{1}{2\,\Delta t}\,\left\langle \left(\Delta z - \langle \Delta z\rangle\right)^2\right\rangle\,,
\end{equation}
where $\langle \ldots \rangle$ indicated an ensemble average. We extracted
$D(z)$ from the trajectory with biasing force
$f=\SI{4}{\kilo\joule\per\mole\per\nano\meter}$ using a time lag of $\Delta t =
\SI{1}{\pico\second}$ and spatially binned the data into 96 bins over
$z\in[-5, 5]\:\si{\nano\meter}$. To obtain a differentiable
diffusion coefficient, we applied a discrete Fourier analysis, for
details see the Supporting Information (section \SImethods).

\subparagraph{Toy Model Simulation and Features}
The overdamped Langevin toy model described in the Results section was simulated
using the Euler--Maruyama scheme with timestep $\Delta t = 10^{-4}$, enforcing
periodicity via $x \leftarrow x \bmod 1$. For driving forces $f \in
\{1,2,4,8,13,23,40\}$ we generated approximately $2.5\times 10^{5}$
statistically independent samples each, using a sampling interval longer than
the autocorrelation time. For network inputs we supplied NESS samples $x_0 \sim
\pness(x)$ together with the corresponding force $f$, restricting positional
Fourier features to the first mode. Although $U(x)$
(Eq.~\eqref{eq:toy_model_sde}) contains two harmonics, excluding higher modes
avoids injecting system-specific prior structure that would not transfer to more
complex applications.

\subparagraph{Neural Network Training}
The diffusion model was implemented via the JAX machine learning
framework,\cite{jax} the Flax neural network
library,\cite{flax} and the optimization framework Optax.\cite{optax} 
An exponential noise schedule was employed, defined by\cite{song2020score}
\begin{equation}
\sigma_\tau = \sigma_\text{min}^{1-\tau} \sigma_\text{max}^\tau \quad\text{and}\quad \alpha_\tau = \frac{1}{2}\partial_\tau (\sigma_\tau^2)\,,
\end{equation}
where $\sigma_{\text{min}}$ and $\sigma_{\text{max}}$ represented the minimum and
maximum noise levels, respectively. For the periodic system under consideration,
$\sigma_\text{max}$ was set to $L/2$, ensuring that the noising reached
the uniform latent potential. For the MD model, a minimum noise level of
$\sigma_\text{min} = 5 \cdot 10^{-4}L$ was chosen to ensure a high spatial
resolution in the learned potential energy function. Since the toy model
exhibited a much simpler energy landscape, we set $\sigma_\text{min} = 10^{-2}L$
to facilitate training.

The diffusion models were trained for $50$ epochs using the AdamW
optimizer\cite{loshchilov17} with a mini-batch size of $512$. The learning rate
schedule comprised an initial linear warm-up phase from $5\times 10^{-7}$ to $5 \times 10^{-3}$
over the first 5 epochs, followed by a cosine decay to $5\times 10^{-7}$ throughout the
remaining epochs.
All available samples with spacing of $\delta t = \SI{0.2}{\pico\second}$ were used for training.

Training an energy-based diffusion model determines the potential energy up to a
diffusion-time-dependent additive constant, $C(\tau)$. To ensure a well-defined
potential energy function, $U^\theta(x_\tau, \tau)$, and promote its smoothness with
respect to the diffusion time, $\tau$, a regularization term was incorporated into
the loss function
\begin{equation}
    \mathcal{L} = \mathbb{E}_{x_0\sim p_0, x_\tau\sim p(x_\tau|x_0)} \left[\left\|\varepsilon^\theta(x_\tau, \tau) - \varepsilon\right\|^2_{\text{pbc}}\right] + \lambda \mathcal{L}_\text{reg}\,,
\end{equation}
where $\left\|\cdot\right\|_{\text{pbc}}$ denoted the periodic norm (i.e., the
Euclidean norm evaluated under periodic boundary conditions), $\lambda =
10^{-5}$ was the regularization strength, and $\mathcal{L}_\text{reg}$ was defined
as
\begin{equation}
    \mathcal{L}_\text{reg} = \mathbb{E}_{x_0\sim p_0, x_\tau\sim p(x_\tau|x_0)} \left[\left\|\partial_\tau U^\theta(x_\tau, \tau)\right\|^2\right]\,.
\end{equation}
This regularization term enhanced the stability and convergence of the training
process.\cite{mate2025solvation}
For a pseudocode of the algorithm we refer to the Supporting Information (algorithm
\SIalgo). For a more detailed theoretical description on how to train a
diffusion model we refer to Song et al.\cite{song2020score}.

The neural network architecture consisted of a fully connected network. For the
lipid system, this network was configured with five hidden layers, each
containing 64 neurons. In contrast, for the toy model, a more compact
architecture was employed, featuring three hidden layers with 32 neurons each.
The adoption of a less complex network for the toy model was justified by the
higher minimal noise level utilized during its training, which relaxed the
demands on model capacity. The Swish activation function\cite{ramachandran17}
was utilized within these layers to ensure differentiability throughout the
network.

\section{Results and Discussion}\label{sec:results}

\begin{figure*}
  \centering
  \includegraphics{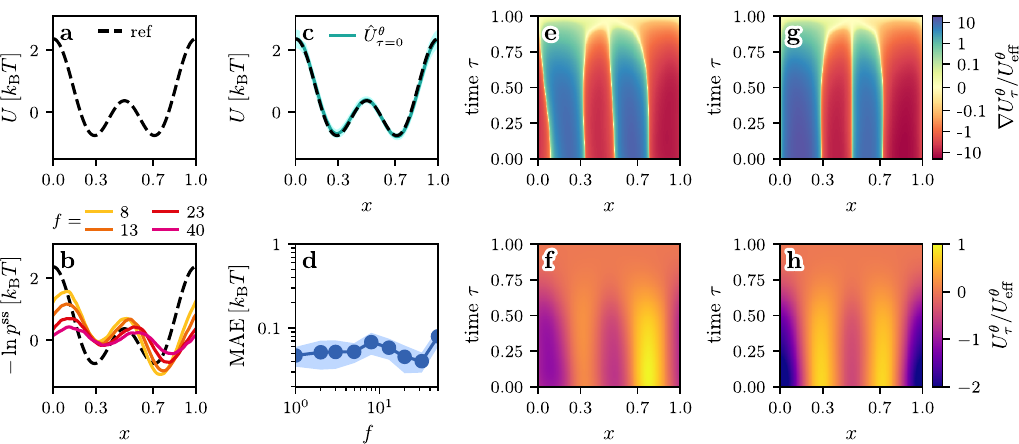}
  \caption{
    Learning the energy landscape of a driven 1D Langevin toy model. 
    (a) Reference free energy profile. 
    (b) Non-equilibrium steady-state distributions $\pness$ of the toy
    model under various constant driving forces $f$. 
    (c) Inferred free energy landscapes $\hat{U}_\theta$ obtained from the
    diffusion model, averaged over twenty independent trials for each biasing
    force; shaded regions denote $\pm$ standard deviation. 
    (d) Mean absolute error (MAE) between the inferred and reference free energy
    profiles as a function of the driving force $f$. Shaded regions denote
    $\pm$ standard deviation over 20 independent trials.
    (e--h) Interpolation between latent space ($\tau=1$) and data space ($\tau=0$) in case
    biasing force $f=13$, illustrating: (e,\,g) Training objective $\nabla U^\theta_\text{eff}(x_\tau,
    \tau)$ and learned equilibrium score $\nabla U^\theta_\tau(x_\tau)$,
    respectively.
    (f,\,h) Learned negative log-density $U^\theta_\text{eff}(x_\tau, \tau)$ and
    $U^\theta_\tau(x_\tau)$, respectively.
    }
    \label{fig:toy_model}
\end{figure*}

\subsection{Enforcing Periodicity in Fokker--Planck Score Learning}

The Fokker--Planck Score Learning framework involves training a diffusion model
with a score function derived from the steady-state solution of the
Fokker--Planck equation for a driven, periodic system.
In Eq.~\eqref{eq:score_with_correction} we established the score function with
the term $\Delta s^\theta(x_\tau, \tau, L)$, defined in
Eq.~\eqref{eq:pbc_correction_term}, arises from the periodicity constraint on
the NESS distribution $\pness(x)$ and ensures that the learned distribution is
periodic.

However, if the neural network parametrizing the potential
$U^\theta_\tau(x_\tau)$ is explicitly constructed to be periodic, such that
$U^\theta_\tau(x_\tau) = U^\theta_\tau(x_\tau + L)$, the integral term within
$\Delta s^\theta(x_\tau, \tau, L)$ (cf. Eq.~\eqref{eq:pbc_correction_term})
becomes a constant with respect to $x_\tau$. Consequently, its gradient, $\Delta
s^\theta(x_\tau, \tau, L)$, vanishes. This simplifies the score function to
\begin{equation}
    s^\theta(x_\tau, \tau, L) = -\beta \nabla U^\theta_\text{eff}(x_\tau, \tau) - \nabla \ln D(x_\tau, \tau)\label{eq:score_no_correction}\,.
\end{equation}
The evaluation of the full correction term $\Delta s^\theta(x_\tau, \tau, L)$
involves an integral in Eq.~\eqref{eq:pbc_correction_term}, which can
be computationally demanding.
In a simple ablation study, we found that using the simplified score
in Eq.~\eqref{eq:score_no_correction} not only yields virtually the same results,
but is also computationally significantly faster and numerically more stable
under strong forces when the periodicity of $U^\theta_\tau(x_\tau)$ is inherently enforced by the
network architecture, see Supporting Information (Fig.~\SIdeltas).

Periodicity of $U^\theta_\tau(x_\tau)$ can be readily enforced by employing a
Fourier series embedding $e$ for the positional features, as proposed by
Tancik \emph{et al.}\cite{tancik20}
\begin{equation}\label{eq:fourier_feature_parametrization}
e(x_\tau, \tau) = \left(\left\{\sin\left(\tfrac{2\pi n}{L}x_\tau\right), \cos\left(\tfrac{2\pi n}{L}x_\tau\right)\right\}_{n},\tau\right)\,
\end{equation}
where only the embedding vector $e(x_\tau, \tau)$ is provided as input to the
neural network, instead of the raw position $x_\tau$ and diffusion time $\tau$
\begin{equation}
U^\theta(x_\tau, \tau)\to U^\theta(e)\,.
\end{equation}
Here, we have used the first $N$ Fourier modes, with $n \in \{1, \ldots, N\}$.
By adopting this Fourier-feature
parametrization, the network inherently learns a periodic function, thereby
obviating the need for the explicit correction term $\Delta s^\theta(x_\tau,
\tau, L)$ in the score function. This approach is employed in the subsequent
sections.

\subsection{Overdamped Langevin Toy Model}

To demonstrate the principles and capabilities of the proposed Fokker--Planck
Score Learning method, we first consider a prototypical non-trivial steady-state
system. This system involves a particle undergoing driven, overdamped Langevin
dynamics in one dimension on the periodic interval $x\in[0,1]$. The equations of
motion are given by
\begin{equation}
    \frac{\textrm{d}x}{\textrm{d}t} = -\frac{\textrm{d}U(x)}{\textrm{d}x} + f + \sqrt{2}\xi(t)\,,
\end{equation}
where $f$ is a constant external driving force, $k_\text{B}T=1$ is assumed,
$\xi(t)$ represents Gaussian white noise characterized by zero mean and
correlation $\langle \xi(t)\xi(t')\rangle = \delta(t-t')$, and
the periodic potential $U(x)$, shown in Fig.~\subref{fig:toy_model}{a}, is
defined as
\begin{equation}
    \label{eq:toy_model_sde}
    U(x) = \cos(2\pi x) + \cos(4\pi x) + \mathcal{C}\,.
\end{equation}
Here, the constant $\mathcal{C}$ is chosen to ensure normalization of the
Boltzmann factor over the unit interval, i.e., $\int_{0}^{1} e^{-U(x)}
\mathrm{d}x = 1$.
Details of the simulation protocol and feature construction are given in the
methods \hyperref[sec:theory]{section}.

We train a neural network to predict the equilibrium energy potential
$\hat{U}^\theta(x, \tau)$, which for $\tau=0$ corresponds to the potential
energy $U(x)$.
As input, the network receives samples from the NESS distribution $\pness(x)$
depicted in Fig.~\subref{fig:toy_model}{b}.
For all selected driving forces, the neural network is able to
accurately reconstruct the potential energy landscape $U(x)$, as shown in
Fig.~\subref{fig:toy_model}{c}. This holds true as long as the steady-state
distribution is distinguishable from a uniform distribution.
It is noteworthy that, in this instance, the inherent symmetry of the potential
is not explicitly enforced during the model training.
This result can be quantified by the mean absolute error (MAE) between the
learned potential $\hat{U}^\theta_{\tau=0}(x)$ and the reference potential
$U(x)$ defined by
\begin{equation}
    \text{MAE} = \left\langle \left|\hat{U}^\theta_{\tau=0}(x) - U(x)\right|\right\rangle\,.
\end{equation}
Since we are estimating relative free energies, the MAE is computed with
the mean-free potentials, i.e., $\langle U(x)\rangle = 0$.
The MAE is shown in Fig.~\subref{fig:toy_model}{d} as a function of the selected
driving force $f$. It is shown that the MAE remains effectively constant across
all selected forces $f$, and yields high accuracy, 
$\textrm{MAE} \approx 0.05\:k_\text{B}T$.

To visualize the training dynamics and the interplay between the Fokker--Planck
solution (Eq.~\eqref{eq:ness_pbc}) and the learning objective,
Fig.~\subref{fig:toy_model}{e,\,g} illustrate the evolution of the effective
potential gradient, $\nabla U^\theta_\text{eff}(x_\tau, \tau)$, and the
effective potential, $U^\theta_\text{eff}(x_\tau, \tau)$, respectively, across
the diffusion time $\tau \in [0, 1]$. Direct parametrization of these
quantities by a neural network corresponds to a standard DDM or an
energy-based DDM, respectively.
Conversely, Fig.~\subref{fig:toy_model}{f,\,h} depict the learned equilibrium
potential gradient, $\nabla U^\theta_\tau(x_\tau)$, and the equilibrium
potential, $U^\theta_\tau(x_\tau)$, as functions of diffusion time.
This visualization demonstrates that employing the Fokker--Planck
score, as defined in Eq.~\eqref{eq:U_eff_dm} and
Eq.~\eqref{eq:score_no_correction}, is equivalent to learning an energy-based
model from equilibrium samples.

\subsection{Lipid Bilayer}
\begin{figure}[tb]
  \centering
  \includegraphics{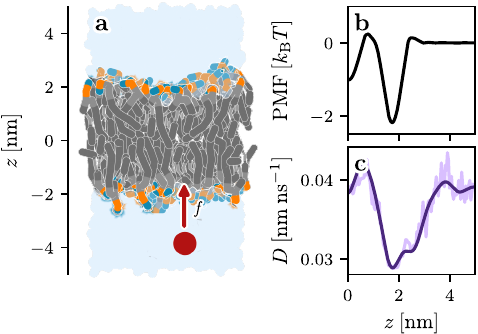}
\caption{
    Lipid bilayer system and its characterization.
    (a) Illustration of the simulated system, depicting a coarse-grained Martini POPC lipid bilayer with a permeating solute (N1 bead type).
    (b) Reference free-energy profile for N1 bead permeation, determined with MBAR via
    high-resolution umbrella sampling with a window spacing of $\delta z =
    \SI{0.02}{\nano\meter}$.
    (c) Position-dependent diffusion coefficient, $D(z)$, of the N1 bead along
    the bilayer normal, derived from molecular dynamics simulations with
    $f=\SI{4}{\kilo\joule\per\mole\per\nano\meter}$.
  }
  \label{fig:lipid_system}
\end{figure}

Having established the principles of Fokker--Planck Score Learning with a toy
model, we now apply this methodology to a more complex and biologically
pertinent system: the lipid bilayer, a fundamental structural component of
biological membranes. Understanding
its free-energy landscape is crucial for elucidating diverse biological
phenomena, including solute permeation, membrane protein function, and lipid
dynamics. Specifically, we use the coarse-grained Martini force field to investigate the permeation of an N1 bead (solute)
through a 1-palmitoyl-2-oleoyl-$sn$-glycero-3-phosphocholine (POPC) lipid bilayer,
depicted in Fig.~\subref{fig:lipid_system}{a}. For more details, a comprehensive
description of the molecular dynamics simulation parameters is presented in the
methods \hyperref[sec:theory]{section}.

Hereafter, $z$ denotes the normal distance of the N1 bead from the bilayer midplane,
which is defined as $z=0$. The regions $z<0$ and $z>0$ correspond to the lower
and upper leaflets, respectively.
Simulations were performed with applied forces of $f\in\{4, 6, 8,
10\}\:\si{\kilo\joule\per\mole\per\nano\meter}$, each for a duration of
$t=\SI{500}{\nano\second}$.
The resulting steady-state distributions are provided in the Supporting
Information (Fig.~\SIlipid{}a).
The applied biasing forces were found to be sufficiently strong to accelerate
the average transition times by up to one order of magnitude, as detailed in the
Supporting Information (Fig.~\SIlipid{}b).

As a reference, we first compute with MBAR the free-energy profile of the N1 bead
by employing umbrella sampling with a high spatial resolution of $\delta z =
\SI{0.02}{\nano\meter}$. The resulting free-energy profile, shown in
Fig.~\subref{fig:lipid_system}{b}, reveals a complex energy landscape with
two minima close to the lipid-water interface and bilayer midplane.
This profile serves as a benchmark for evaluating the performance of our
Fokker--Planck Score Learning method.

In contrast to the toy model, considering biomolecular systems, we expect
in general that the diffusion coefficient is not constant, but rather
position-dependent. As detailed in the \hyperref[sec:theory]{method section}, to
account for this in the Fokker-Planck ansatz, we
need to know the position-dependent diffusion coefficient $D(z)$, which we
estimate from the MD simulations with the weakest biasing force
$f=\SI{4}{\kilo\joule\per\mole\per\nano\meter}$.
Notably, applying a constant force over only $\SI{50}{\nano\second}$ of
MD simulation time suffices to obtain a robust estimate of the
position-dependent diffusion coefficient, which holds for all studied forces
(see Supporting Information Fig.~\SIlipid{}c,d).
For all subsequent training runs we employ the diffusion coefficient
computed from the full trajectory, as shown in
Fig.~\subref{fig:lipid_system}{c}.
In contrast to all-atom MD
simulations, where the diffusion coefficient varies by almost an order of
magnitude,\cite{carpenter2014method} the coarse-grained Martini 3 model yields smaller variations, but remains position dependent. Furthermore, Carpenter \emph{et al.} have shown that the position-dependent diffusion coefficient varies weakly with the
compound being simulated, within reasonable limits of the solute chemistry
\cite{carpenter2014method}. This observation can be used to make repeated use of the diffusion coefficient across a high-throughput screening of compounds at a coarse-grained resolution.\cite{menichetti2019drug}

For the lipid bilayer system, we introduce two refinements to the methodology
previously applied to the toy model.
First, we constrain the neural network to learn only even periodic functions,
i.e., $U^\theta(x) = U^\theta(-x)$. This is achieved by employing exclusively
cosine Fourier modes in the network parametrization. Such symmetry is a common
assumption in lipid bilayer simulations and is often exploited in umbrella
sampling by restricting sampling windows to one half of the periodic domain.
This constraint is physically justified by the expected symmetry of the
potential of mean force (PMF) with respect to the bilayer center.
Second, whereas individual networks were trained for each distinct force in the
toy model, we now train a single network using data aggregated from simulations
performed under multiple driving forces. This approach leverages the fact that
all non-equilibrium simulations, irrespective of the applied force, probe the
same underlying PMF. The benefit of this is that, while weak forces sample the
PMF with high resolution, stronger forces allow for faster sampling of the NESS
distribution and therefore provide complementary information.

For the Fokker--Planck Score Learning approach, the initial 32 Fourier modes
were employed as input features to the neural network to enhance overall
performance, in accordance with the parameterization detailed in
Eq.~\eqref{eq:fourier_feature_parametrization}. To evaluate the efficacy of
this method, the neural network was trained using datasets corresponding to
varying total trajectory lengths, $t \in [12.8, 53.2]\:\si{\nano\second}$.
These datasets were composed of segments randomly selected from each
constant-force simulation. This approach allowed for an assessment of the
method's performance and uncertainty as a function of the available data. For
comparison with conventional umbrella sampling--based MBAR, simulations were
conducted with
a window spacing of $\delta z = \SI{0.1}{\nano\meter}$ over the range $z \in [0,
4.8]\:\si{\nano\meter}$. Each window was simulated for
$t=\SI{500}{\nano\second}$, resulting in a total simulation time of
\SI{24.5}{\micro\second}. To estimate the uncertainty in the umbrella
sampling--based results, both the weighted histogram analysis method (WHAM) and
multistate Bennett acceptance ratio (MBAR) were applied to trajectory
segments of lengths $t \in [1, 100]\:\si{\nano\second}$ from
each window. Since MBAR and WHAM yielded virtually identical results, only the MBAR
results are presented in the following.

\begin{figure}[tb]
  \centering
  \includegraphics{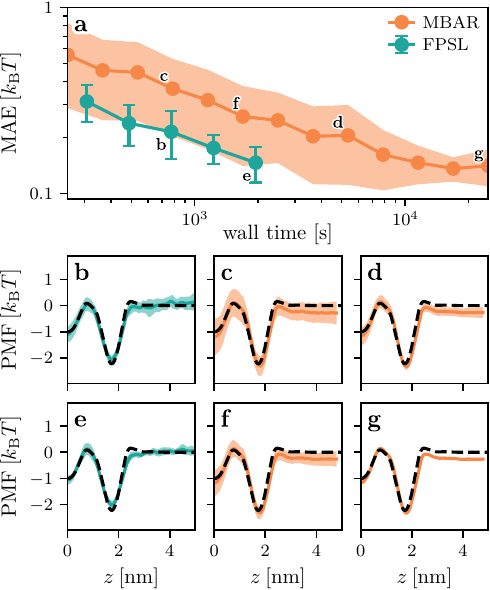}
\caption{
    Performance evaluation of Fokker--Planck Score Learning for reconstructing
    the free-energy profile of N1 bead permeation through a POPC lipid bilayer.
    (a) Mean absolute error (MAE) versus computational wall time for
    Fokker--Planck Score Learning (green) and umbrella sampling with MBAR (orange).
    (b--d) Comparison of reconstructed free-energy profiles. (b) Fokker--Planck
    Score Learning using approximately $\SI{128.9}{\nano\second}$ of MD
    simulation data. (c) MBAR with comparable wall time to (b). (d)
    MBAR with a comparable MAE to (b).
    (e--g) Similar comparison as (b--d), but with Fokker--Planck Score Learning
    utilizing approximately $\SI{325}{\nano\second}$ of MD simulation data. (e)
    Fokker--Planck Score Learning. (f) MBAR with comparable wall
    time to (e). (g) MBAR with a comparable MAE to (e).
    Shaded regions indicate $\pm$ standard deviations.
  }
  \label{fig:mae_results}
\end{figure}

To assess the computational efficiency of Fokker--Planck Score Learning relative
to conventional umbrella sampling--based MBAR, we compare the total computational time
invested. For umbrella sampling, this is defined as the cumulative production
run time across all sampling windows. For our method, it comprises the time for
constant-force pulling MD simulations plus the neural network training duration.
We neglect the time required for the initial equilibration and minimization of the
MD simulation and the time for the MBAR analysis, as these are expected to be
small compared to the total simulation time.

As reference, all simulations and analysis were performed on an Intel\textregistered{}
Core\texttrademark{} i7-12700 CPU and an Nvidia GeForce RTX\texttrademark{} 4060
GPU, yielding a throughput of approximately \SI{18}{\micro\second} of MD
simulation per day for the lipid bilayer system. The training duration for the
diffusion model exhibits a linear scaling with the used simulation data time. On
the aforementioned hardware, the neural network training was found to be
approximately $5.0$ times
faster than generating an equivalent duration of MD simulation data.
Figure~\subref{fig:mae_results}{a} presents the mean absolute error (MAE) of the
reconstructed free-energy profiles as a function of the needed wall time
for both Fokker--Planck Score Learning and umbrella sampling. This comparison
highlights that our method requires around $\SI{2000}{\second}$ to achieve an
error of $\text{MAE}\approx 0.15\:k_\text{B}T$, while MBAR needs
roughly one order of magnitude more time to reach a similar level of accuracy.

To visualize the meaning, we show the mean and variance of the reconstructed
free-energy profile in of Fokker--Planck Score Learning using $\SI{800}{\second}$
wall time, the corresponding MBAR with a comparable wall time, and
the MBAR with a comparable MAE in Fig.~\subref{fig:mae_results}{b--d}, respectively.
It is clearly visible that the Fokker--Planck Score Learning method has no
systematic bias and the largest uncertainty is observed in the water $z\ge
\SI{3}{\nano\meter}$ region. This is expected, as this region is sampled
least in the NESS distribution. In contrast, Fig.~\subref{fig:mae_results}{c}
shows that the MBAR result using the same computational time exhibits an almost
uniform, but much larger, uncertainty across the entire range.
Comparing the MBAR run with a comparable $\text{MAE}$ in
Fig.~\subref{fig:mae_results}{d}, we observe that the uncertainty inside the
membrane is comparable to the Fokker--Planck Score Learning result.
However, we find for MBAR a systematic bias in the free-energy due to the
limited spatial resolution.

Similarly, we compare the results of Fokker--Planck Score Learning using a
longer wall time of approximately $\SI{2000}{\second}$ with MBAR of
similar computational time and MBAR with a comparable MAE in
Fig.~\subref{fig:mae_results}{e--g}. Again, we observe that the Fokker--Planck
Score Learning method yields a free-energy profile with only minor systematic
bias, primarily underestimating the second barrier at
$z\approx\SI{2.2}{\nano\meter}$, and exhibits a reduced uncertainty in the lipid
region. In contrast, due to the limited spatial resolution, we again find that
MBAR slightly under- or overestimates all regions, introducing a systematic bias
into the free-energy profile.
The results indicate that our method not only achieves a lower MAE
at a given computational cost compared to MBAR but also exhibits
significantly less systematic bias. This is of particular
importance for applications in drug discovery and membrane-protein function.
Using WHAM in place of MBAR yields virtually identical results. An illustrative
comparison of the individual PMF reconstructions from MBAR and WHAM is shown in
the Supporting Information (Fig.~\SIumbrella{}).

Finally, we want to discuss the robustness of our method with respect to the
choice of the hyperparameters, in particular the number of Fourier features and
the minimal noise level $\sigma_\text{min}$.
While we find that for a reasonable range of Fourier features $N_f \in [1, 32]$
the results are robust with only a minor decrease in the MAE with increasing
$N_f$, we observe that for greater values of $N_f$ the training becomes less
stable, leading to a slight increase in the MAE, as shown in the Supporting
Information (Fig.~\SIhyper{}b). The choice of the minimal noise level
$\sigma_\text{min}$ is more critical, as it directly controls the spatial
resolution of the learned potential. We find that for
$\sigma_\text{min}\in[10^{-4}L, 3 \times 10^{-3}L]$ the results are robust,
while for larger values the MAE increases due to insufficient spatial resolution
and for smaller values the training becomes unstable due to overfitting, as
shown in the Supporting Information (Fig.~\SIhyper{}a).

The application of the Martini 3 coarse-grained force field
substantially reduces simulation times, enhancing the computational
feasibility of methods such as umbrella sampling--based MBAR for the systems
investigated. For all-atom molecular dynamics simulations, which are
significantly more computationally intensive, the neural network training
duration is anticipated to constitute a minor fraction of the total simulation
time. This scenario would permit more elaborate optimization of the neural
network architecture and training parameters, potentially further improving the
convergence and precision of the Fokker--Planck Score Learning method presented
herein.

\section{Conclusion}

We have introduced Fokker--Planck Score Learning, a physics-informed,
score-based diffusion framework to efficiently reconstruct free-energy profiles
under periodic boundary conditions (PBCs) and non-equilibrium pulling. By
mapping a PBC simulation onto a Brownian particle in a periodic potential, we
make use of the steady-state solution of the Fokker--Planck equation to inform
the score of a denoising diffusion model---the Fokker--Planck score. Training a
neural network on non-equilibrium trajectory data yields a
diffusion-time-dependent potential whose zero-time limit recovers the underlying
potential of mean force (PMF).

A key advantage of our approach is the physics-based prior: the analytic
non-equilibrium steady-state (NESS) solution for periodic systems constrains the
model to a limited space of physically plausible probability densities,
dramatically reducing the amount of sampling needed. We demonstrated on a
one-dimensional toy model that the learned potential is recovered to high
accuracy---including for large driving forces. In a complex biomolecular 
application (solute permeation
through a POPC bilayer modeled at a coarse-grained level), our method achieved
up to one order-of-magnitude improvements in efficiency over MBAR with umbrella sampling,
while systematically producing lower variance.

Because we parameterize the network to be explicitly periodic (via Fourier
features), the correction term $\Delta s^\theta(x_\tau, \tau, L)$
(Eq.~\ref{eq:pbc_correction_term}) in the NESS score vanishes---simplifying both
implementation and training. In particular, the integral present in $\Delta
s^\theta(x_\tau, \tau, L)$ incurs a significant computational footprint, but can
simply be omitted by informing $U^\theta(x_\tau,\tau)$ of its periodicity.

Looking ahead, we expect even greater gains in atomistic simulations, where
generating sufficient samples takes significantly more compute time; the
diffusion-model training then becomes a minor overhead that could be further
optimized. Extending Fokker--Planck Score Learning to higher-dimensional PMFs is
straightforward in principle, but will require careful strategies for sampling
and modeling the orthogonal degrees of freedom. We anticipate that coupling our
score-based approach with enhanced sampling along multiple coordinates will
unlock efficient free-energy estimation across a broad range of complex
molecular systems.

\section*{Acknowledgments}
We thank Luis J. Walter, Luis Itzá Vázquez-Salazar, and Sander Hummerich
for their valuable feedback on the manuscript, and the anonymous reviewers for their
constructive comments that helped improve this work.

We acknowledge support by the Deutsche Forschungsgemeinschaft (DFG, German Research Foundation) under Germany's Excellence Strategy EXC 2181/1 - 390900948 (the Heidelberg STRUCTURES Excellence Cluster) and Heidelberg University through the Research Council of the Field of Focus 2 ``Patterns and Structures in Mathematics, Data, and the Material World.''

\section*{Data availability statement}
A Python package implementing the Fokker--Planck Score Learning framework is
available under an open-source license at
\url{https://github.com/BereauLab/fokker-planck-score-learning}. All
constant-force simulation data have been deposited in the NOMAD repository and
can be accessed via
\href{https://dx.doi.org/10.17172/NOMAD/2025.06.26-1}{DOI:10.17172/NOMAD/2025.06.26-1}.

\bibliography{lit_iso4}

\end{document}


\title{Supporting Information for "Fokker--Planck Score Learning: Efficient Free-Energy Estimation under Periodic Boundary Conditions"}

\author{Daniel Nagel}
\affiliation{%
 Institute for Theoretical Physics, Heidelberg University, 69120 Heidelberg, Germany
}%
\author{Tristan Bereau}%
\email{bereau@uni-heidelberg.de}
\affiliation{%
 Institute for Theoretical Physics, Heidelberg University, 69120 Heidelberg, Germany
}%
\affiliation{%
 Interdisciplinary Center for Scientific Computing (IWR), Heidelberg University, 69120 Heidelberg, Germany
}%

\date{1 October 2025}

\maketitle

\section{Deriving Probability Distributions\label{sec:si_deriving_prob_dist}}

In this section, we examine the derivation of probability distributions for
stochastic processes. We begin with the Langevin equation and proceed to derive
the Fokker--Planck equation. Subsequently, we derive the stationary distribution
of the Fokker--Planck equation. For a comprehensive overview of the
Fokker--Planck equation, we refer to the literature\cite{risken1996fokker}.

Consider an overdamped Langevin equation of the form
\begin{equation}
\mathrm{d}x = \frac{1}{\gamma(x)} (-\nabla_x U_\text{eff}(x, t))\mathrm{d}t + \sqrt{\frac{2}{\beta \gamma(x)}} \mathrm{d}W\;,
\end{equation}
where $x$ denotes the position of the particle, $\gamma(x)$ represents the
position-dependent friction coefficient, $U_\text{eff}(x, t) = U(x, t) - fx$ is
the effective potential, $\beta$ is the inverse temperature, and $W$ is the
Wiener process. Utilizing Einstein's relation $D(x)\gamma(x) = \beta^{-1}$, the
Langevin equation can be expressed as
\begin{equation}
\mathrm{d}x = -\beta D(x)\nabla_x U_\text{eff}(x, t)\mathrm{d}t + \sqrt{2D(x)} \mathrm{d}W\;.
\end{equation}
The Fokker--Planck equation for the probability distribution $p_t(x)$ is given
by
\begin{align}
\partial_t p_t(x) &= \nabla_x\left[\beta D(x)\nabla_x U_\text{eff}(x,t) p_t(x)\right] + \nabla_x^2 \left[D(x) p_t(x)\right]\nonumber\\
&= -\nabla_x J_t(x)\;,
\end{align}
where
\begin{align}
J_t(x) = -D(x) \Big[&
    \beta \nabla_x U_\text{eff}(x,t) p_t(x) \nonumber\\
    &+ \nabla_x \ln D(x) p_t(x) + \nabla_x p_t(x)\Big]\;.
\end{align}
For a steady-state distribution $p_t(x) = p^\text{ss}(x)$, where $\partial_t
p_t(x) = 0$, the current $J_t(x)$ remains constant. Introducing the auxiliary
variable $\kappa(x,t) = D(x) e^{\beta U_\text{eff}(x, t)}$, we obtain
\begin{equation}
\nabla_x (\kappa(x,t) p^\text{ss}(x)) = - \frac{J}{D(x)} \kappa(x,t)\;.
\end{equation}
Integrating this equation from $x\to x+L$ leads to
\begin{equation}
p^\text{ss}(x) = \frac{J}{1 - e^{-\beta f L}}\frac{1}{D(x)} e^{-\beta U_\mathrm{eff}(x, t)} \int_x^{x+L}\mathrm{d}y\, e^{\beta U_\mathrm{eff}(y,t)}\;,
\end{equation}
where we used $\kappa(x+L, t) = e^{-\beta f (1-t) L} \mu(x, t)$, $D(x+L)=D(x)$, and $U_t(x+L)=U_t(x)$. From the normalization of the steady-state probability $\int_0^L \mathrm{d}x\:p^\text{ss}(x)=1$ it follows directly for the flux
\begin{equation}
J = (1 - e^{-\beta fL}) \left[\int_0^L \mathrm{d}x\: \frac{1}{D(x)} e^{-\beta U_\text{eff}(x, t)}\int_{x}^{x+L} \mathrm{d}y\: e^{\beta U_\text{eff}(y, t)}\right]^{-1}\;.
\end{equation}
It should be noted that for the diffusion time $\tau$, the prefactor  of $p^\text{ss}$ and $J$ takes the following form $1-e^{-\beta fL} \to 1 - e^{-\beta (1-\tau) fL}$.
Since only the gradient of the probability distribution is needed for training a diffusion model, we neglect the integration constant which yields
\begin{equation}
p_t(x) \propto \frac{1}{D(x)} e^{-\beta U_\mathrm{eff}(x, t)} \int_x^{x+L}\mathrm{d}y\, e^{\beta U_\mathrm{eff}(y,t)}\;.
\end{equation}

\section{Methods}

\subsection{Discrete Fourier Analysis of the Diffusion Coefficient}
To obtain a differentiable diffusion coefficient $D(z)$ from the trajectory with
\begin{equation}
D_N(z) = D_0 + \sum_{n=1}^{N}\alpha_n \cos(\tfrac{2\pi n}{L} z),
\end{equation}
where the coefficients $D_0$ and $\alpha_n$ were given by
\begin{align}
D_0 &= \int_0^L \mathrm{d}z\, D(z), \quad\text{and}\\
\alpha_n &= 2 \int_0^L \mathrm{d}z\, \cos(\tfrac{2\pi n}{L} z) D(z)\,.
\end{align}

\FloatBarrier
\subsection{Training Details of the Neural Network}
{\linespread{1.0} 
\begin{algorithm}
\caption{
  One training iteration of Fokker--Planck Score Learning.
  For the sake of simplicity, we omitted the position-dependent diffusion
  coefficient $D(x)$ term and the integral based term $\Delta s^\theta$ in the
  prediction $\hat{\varepsilon}^\theta$ (indicated with $(*)$).
}\label{alg:fpsl}
\DontPrintSemicolon
\SetKwInOut{Input}{Input}\SetKwInOut{Output}{Output}
\Input{Sample size $N$; Batch size $B$; NESS trajectories; period $L$; external force $f$; noise schedule $\sigma_\tau$; Fourier feature size $N_f$;}
\Output{Updated parameters $\theta$}
\For{$B\in N$}{
\For{$i\in B$}{
    Wrap $x_0^{(i)} \gets x_0^{(i)} \bmod L$;\;
    Draw $\tau^{(i)} \sim \mathcal{U}(0,1)$ and $\varepsilon^{(i)} \sim \mathcal{N}(0,1)$;\;
    Sample $x_\tau^{(i)} \gets x_0^{(i)} + \sigma_{\tau^{(i)}} \varepsilon^{(i)} \bmod L$;\;
    Embed $e_\tau^{(i)} \gets \left(\left\{\sin\left(\tfrac{2\pi n}{L}x_\tau^{(i)}\right), \cos\left(\tfrac{2\pi n}{L}x_\tau^{(i)}\right)\right\}_{n=1}^{N_f},\tau^{(i)}\right)$;\;
    Evaluate and take gradient of $\hat{\varepsilon}^{(i)} \gets - (1-\tau^{(i)}) \left[\epsilon^\theta(e_\tau^{(i)}) - \sigma_{\tau^{(i)}} f x_\tau^{(i)}\right] + (*)$;\;
}
Update $\theta$ with optimizer (e.g. AdamW) on $\mathcal{L} = \sum_i \mathcal{L}_i/B$ (see main text);\;
}
\end{algorithm}
}


\FloatBarrier
\section{Results}
\begin{figure}
  \centering
  \includegraphics{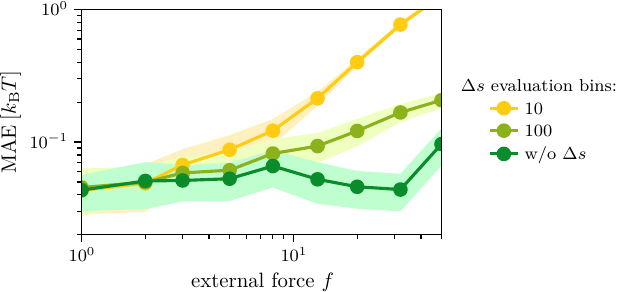}
  \caption{
    Ablation study of the toy-model system (see main text) evaluating the periodic
    correction term $\Delta s^\theta(x_\tau, \tau, L) = \nabla \ln
    \int_{x_\tau}^{x_\tau + L} \mathrm{d}y\: e^{\beta U^\theta_\text{eff}(y,
    \tau)}$ when enforcing periodicity via Fourier features.
    The integral is approximated on a grid with 10 or 100 points, or omitted
    entirely. For small forces, all results agree well. However, with increasing
    force, in $U^\theta_\text{eff}(x, \tau) = (1-\tau)[U(x) - f x] + \tau \ln L$,
    the tilting term $f x$ dominates and the integral becomes numerically
    challenging to optimize for.
  }
\end{figure}

\begin{figure}
    \centering
    \includegraphics[width=\linewidth]{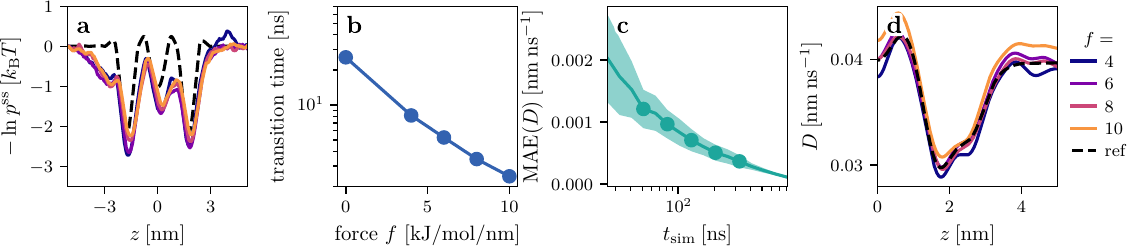}
    \caption{
        POPC lipid bilayer system.
        (a) Steady-state probability distribution, $p^\text{ss}(z)$,
        corresponding to the center-of-mass (COM) distance $z$ of a POPC lipid
        bilayer relative to the N1 bead, under various constant pulling forces
        $f \in \{4, 6, 8, 10\}\:\si{\kilo\joule\per\mole\per\nano\meter}$. The
        black dashed line indicates the reference Potential of Mean Force (PMF)
        obtained via umbrella sampling; further details are available in the
        main text.
        (b) Estimated mean transition time from the central barrier region
        (around $z=0$) to an adjacent potential well. The application of pulling
        force reduces this transition time by more than an order of magnitude.
        (c) Mean absolute error of the position-dependent diffusion coefficient
        $D(z)$ estimated from MD simulations of duration $t_\text{sim}$ under an
        applied force $f = \SI{4}{\kilo\joule\per\mole\per\nano\meter}$.
        Reference values were computed using the full simulation length,
        $t_\text{sim} = \SI{5}{\micro\second}$. Circles denote the results
        presented in the main text, see Fig.~4.
        (d) Position-dependent diffusion coefficient, $D(z)$, estimated from the
        steady-state probability distribution $p^\text{ss}(z)$ (as depicted in
        panel a), and contrasted with values obtained from a reference
        equilibrium simulation (indicated by the black dashed line).
    }
\end{figure}

\begin{figure}
  \centering
  \includegraphics[width=0.95\linewidth]{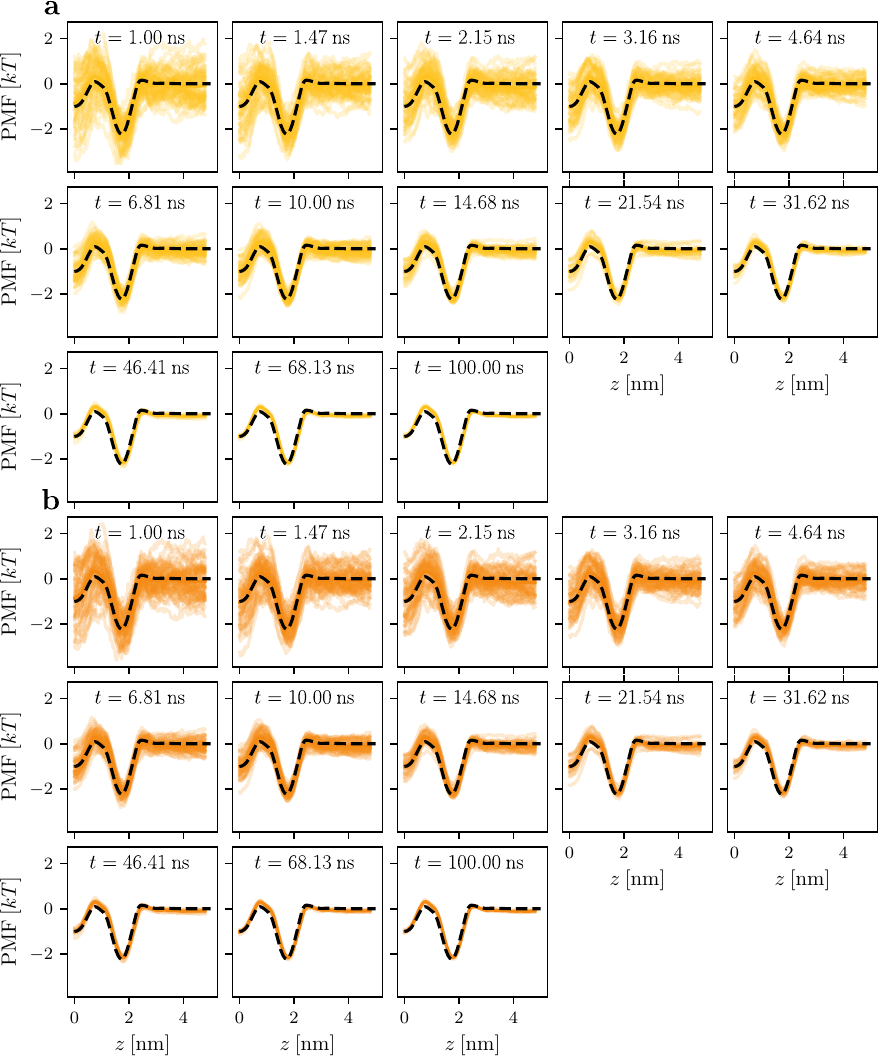}
  \caption{
    The potential of mean force (PMF) for a POPC lipid bilayer, as determined
    based on umbrella sampling\cite{torrie1977nonphysical}.
    A total of 49 simulation windows were utilized, spanning the range $z \in
    [0, 4.8]$, with a spacing of $\SI{0.1}{\nano\meter}$ between adjacent
    windows.
    For each panel, the PMF was estimated using (a, yellow) the weighted histogram analysis
    method (WHAM)\cite{kumar92, hub10} or (b, orange) the multistate Bennett
    acceptance ratio (MBAR)\cite{shirts2008}, incorporating the MD
    simulation time per window as indicated.
    To assess statistical uncertainty, the analysis was replicated up to 20
    times.
    }
  \label{fig:umbrella_sampling}
\end{figure}

\begin{figure}
  \centering
  \includegraphics{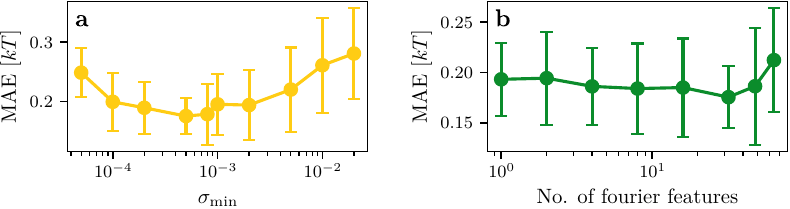}
  \caption{
    Hyperparameter optimization for the Fokker--Planck Score Learning (FPSL)
    method on a POPC lipid bilayer system.
    Mean absolute error (MAE) of the potential of mean force (PMF)
    with respect to reference values obtained from umbrella sampling/MBAR, as a
    function of (a) the minimal noise level $\sigma_\text{min}$ used in the
    noise schedule and (b) the number of Fourier modes used in the neural network
    architecture.
    For all runs shown here, we used a total MD simulation time of
    \SI{204.8}{\nano\second} and the hyperparameters reported in the main paper.
    }
  \label{fig:hyperparameter_optimization}
\end{figure}

\FloatBarrier
\bibliography{lit_iso4}